\documentclass[pdflatex,sn-basic]{sn-jnl}

\usepackage{graphicx}%
\usepackage{multirow}%
\usepackage{amsmath,amssymb,amsfonts}%
\usepackage{amsthm}%
\usepackage{mathrsfs}%
\usepackage[title]{appendix}%
\usepackage{xcolor}%
\usepackage{textcomp}%
\usepackage{manyfoot}%
\usepackage{booktabs}%
\usepackage{algorithm}%
\usepackage{algorithmicx}%
\usepackage{algpseudocode}%
\usepackage{listings}%
\usepackage{placeins}%

\theoremstyle{thmstyleone}%
\theoremstyle{thmstyletwo}%

\theoremstyle{thmstylethree}%

\begin{document}

\title[Cross Subtype Transferability in Seyfert AGN]
{Cross Subtype Transferability of Machine Learning Photometric Redshift Relations in Low Redshift Seyfert AGN}


\author*[1]{\fnm{Uzay} \sur{Aydin}}\email{uzayaydiin@gmail.com}

\affil*[1]{
\orgdiv{Department of Astronomy and Space Sciences},
\orgname{Erciyes University},
\orgaddress{
\city{Kayseri},
\postcode{38039},
\country{Türkiye}
}
}


\abstract{
Photometric redshift estimation for active galactic nuclei (AGN) is complicated by the combined effects of host-galaxy light, nuclear emission, dust attenuation, and broadband spectral diversity. We investigate whether machine-learning photo-z relations trained on one low-redshift Seyfert subtype remain valid when transferred to another, and whether probabilistic subtype classification can be used to identify sources for which a specialised regressor is reliable. Using spectroscopically selected Seyfert~I and Seyfert~II samples from SDSS, matched to AllWISE photometry over $0 < z_{\rm spec} \leq 0.6$, we constructed a common 45-feature representation from SDSS $ugriz$ and WISE W1--W4 data. Random Forest and XGBoost regressors were evaluated within each subtype, followed by controlled cross-subtype transfer tests, redshift- and sample-size-matched experiments, feature ablations, and an independent classifier-gated regression test.

The subtype-specific models achieved strong within-sample performance, with the Seyfert~II model reaching $R^2 = 0.965$ and $\sigma_{\rm NMAD} = 0.0169$. However, transfer between Seyfert~I and Seyfert~II produced a clear and asymmetric degradation in accuracy that persisted after matching the samples and restricting the photometric inputs. A probabilistic Seyfert classifier further identified subsets for which the Seyfert~II regressor was more reliable, while extrapolation beyond the redshift range represented in training produced systematic underestimation. These results demonstrate that AGN photo-z performance depends strongly on the population and redshift domain represented in the training data, supporting subtype-aware calibration and applicability-based source selection.
}




\maketitle
\section{Introduction}

Spectroscopic redshifts provide the most reliable distance information for galaxies and active galactic nuclei (AGN), but spectroscopy cannot be obtained for every source detected by modern wide field imaging surveys. This limitation is increasingly important as surveys such as the Vera C. Rubin Observatory Legacy Survey of Space and Time (LSST) expand the number of photometrically detected galaxies far beyond the number with spectroscopic follow up \citep{York2000,Ivezic2019}. Photometric redshift estimation, hereafter photo-\(z\), therefore remains an essential tool for large survey analyses \citep{Baum1962,Koo1985,Connolly1995,Csabai2000}.

Photo-\(z\) methods are commonly divided into template based and empirical approaches. Template based methods compare observed photometry with theoretical or empirical spectral energy distributions (SEDs), whereas machine learning methods infer the relation between photometric observables and redshift directly from representative training data \citep{Bolzonella2000,CarrascoKind2013,Salvato2019}. AGN are more difficult than normal galaxies because their broadband photometry can contain varying contributions from the host galaxy, the nuclear continuum, emission lines, and dust related infrared emission. These components can broaden or distort the colour redshift relation and increase degeneracies and large redshift errors \citep{Richards2001,Salvato2009,Salvato2011}.

AGN photo-\(z\) studies have consequently developed methods that explicitly account for population diversity. Early SDSS quasar studies recovered useful redshift information from broadband colours and empirical colour redshift relations \citep{Richards2001,Weinstein2004}. Later work introduced probabilistic approaches, AGN and host galaxy template combinations, and classification assisted fitting in which sources are first assigned to a more appropriate spectral class before redshift estimation \citep{Bovy2012,Assef2010,Salvato2009,Salvato2011,FotopoulouPaltani2018}. Empirical methods based on Random Forests, prediction trees, boosted ensembles, and deep learning have also been applied to AGN and quasar samples over a wide range of survey depths and redshifts \citep{CarrascoKind2013,Nakoneczny2019,Nakoneczny2021,PasquetItam2018,DIsanto2018,Zhu2025,Roster2024}.

A related strategy is to divide a heterogeneous source population before regression. For quasars, a two stage approach in which sources are first assigned to redshift intervals and then passed to regressors specific to those intervals has been shown to improve photo-\(z\) performance \citep{Li2022}. AGN studies in deep radio continuum fields have likewise shown that no single method necessarily performs best across samples selected at different wavelengths \citep{Duncan2018}. These results motivate a direct test of whether a regression relation calibrated for one physically distinct AGN population can be transferred to another.

Seyfert galaxies provide a useful low-redshift setting for this question. In broad-line Type I AGN, the unobscured nuclear continuum can contribute strongly to the optical SED, while in narrow-line Type II AGN the optical continuum is generally more strongly influenced by the host galaxy because the central engine is obscured. The relative contributions of host and nucleus therefore differ systematically between the two populations \citep{Kewley2001,Kauffmann2003}. Host dominated, low luminosity AGN have also been reported to reach photo-\(z\) quality closer to that of inactive galaxies than samples dominated by nuclear emission \citep{Brescia2021}.

Machine learning has previously been used to distinguish broad-line and narrow-line AGN photometrically. \citet{Cavuoti2014} addressed emission line galaxy classification, while \citet{Zeraatgari2024} extended photometric classification to a broader multi class scheme. \citet{Mechbal2024} trained a Random Forest regressor jointly on a mixed sample of Type I and Type II AGN while predicting subtype as a separate output. These studies did not directly evaluate the two directional transfer problem considered here: whether a photo-\(z\) relation trained on one Seyfert subtype retains its accuracy when applied to the other, and whether subtype classification can be used to decide when a specialised regressor should be applied.

We investigate this question within the common low-redshift interval
\[
0 < z_{\mathrm{spec}} \leq 0.6.
\]
The interval provides substantial spectroscopic samples of both Seyfert subtypes while limiting the analysis to a shared redshift domain, thereby reducing the need for extrapolation beyond the spectroscopic support of the training data. It also allows the transfer tests to be interpreted within a controlled low-redshift regime rather than across strongly different redshift ranges.

The study has three linked aims. First, we construct subtype specific photo-\(z\) models from SDSS optical and WISE mid-infrared photometry using a common 45 feature representation. Second, we perform controlled Random Forest experiments in both transfer directions and repeat the comparison with matched sample sizes and spectroscopic redshift distributions. We further test whether the transfer behaviour persists when the feature representation is restricted by photometric family or wavelength range. Third, we evaluate a probabilistic Seyfert subtype classifier as a selection step before applying the Seyfert~II regressor, including an external test on luminous QSOs that are excluded from classifier training and threshold selection.

The purpose is not to develop or claim a general-purpose AGN photo-\(z\) estimator for unknown survey sources. Rather, we use spectroscopically defined Seyfert subtypes as controlled populations to test how empirical machine-learning photo-\(z\) relations transfer across population boundaries, and whether probabilistic subtype selection can restrict a specialised regressor to sources that are better represented by its training domain.

\section{Data}

The analysis is based on three spectroscopically defined samples:
a narrow-line Seyfert~II sample selected from SDSS DR19, a
low luminosity broad-line Seyfert~I candidate sample selected from
SDSS DR16Q, and a luminous QSO comparison sample derived from the
same broad-line catalogue. The optical samples were matched
with the AllWISE Source Catalogue, and a common photometric
representation was constructed from complete SDSS \(ugriz\) and
WISE \(W1\)--\(W4\) measurements. The sample selection criteria,
cross matching procedure, and photometric feature construction are
described below.  

\subsection{Optical Sample Selection from SDSS}
\label{sec:optical_sample_selection}

We constructed two spectroscopically defined optical samples from SDSS: a narrow-line Seyfert~II sample and a broad-line Type~I AGN sample. Different SDSS data products were used because the spectroscopic measurements required to define these populations are not provided in a single catalogue.

The Seyfert~II sample was selected from SDSS Data Release~19
\citep{AdamanePallathadka2025}. The parent catalogue was queried from the SDSS database through the CasJobs interface using Structured Query Language (SQL). We first restricted the parent sample to spectroscopically
observed sources classified by the SDSS pipeline as
\texttt{class = GALAXY} and \texttt{subClass = AGN}. This
initial selection adopts the internal SDSS spectroscopic
classification scheme to identify galaxies whose narrow
emission line ratios are consistent with AGN excitation
\citep{Bolton2012}. The operational criterion used for this
classification is

\begin{equation}
\log_{10}
\left(
\frac{[\mathrm{O\,III}]\lambda5007}{\mathrm{H}\beta}
\right)
>
0.7
-
1.2
\left[
\log_{10}
\left(
\frac{[\mathrm{N\,II}]\lambda6584}{\mathrm{H}\alpha}
\right)
+
0.4
\right].
\end{equation}

We subsequently further restricted this parent sample using the
$[\mathrm{N\,II}]\lambda6584/\mathrm{H}\alpha$
Baldwin--Phillips--Terlevich (BPT) diagnostic diagram
\citep{Baldwin1981}. Sources were required to lie above the
theoretical maximum starburst demarcation of \citet{Kewley2001}:

\begin{equation}
\log_{10}\left(
\frac{[\mathrm{O\,III}]\lambda5007}{\mathrm{H}\beta}
\right)
>
\frac{0.61}{
\log_{10}\left(
\frac{[\mathrm{N\,II}]\lambda6584}{\mathrm{H}\alpha}
\right)-0.47
}
+1.19.
\label{eq:kewley_bpt}
\end{equation}

The sample was restricted to
\(0 < z_{\mathrm{spec}} \leq 0.6\)
and to sources with valid extinction corrected SDSS photometry in the \(u\), \(g\), \(r\), \(i\), and \(z\) bands. These criteria produced a sample of 23,797 Seyfert~II galaxies. The spectroscopic redshift was retained only as the reference quantity for model training and evaluation and was not included among the photometric input features.

The broad-line Type~I sample was constructed from the SDSS DR16 Quasar Catalogue \citep{Lyke2020}. We first selected 31,562 sources within
\(0 < z_{\mathrm{spec}} \leq 0.6\)
and matched them to the DR16Q quasar property catalogue of
\citet{WuShen2022}. This catalogue provides measurements of the continuum and broad emission lines, together with improved systemic redshift estimates.

\begin{figure*}
    \centering
    \includegraphics[width=0.92\textwidth]{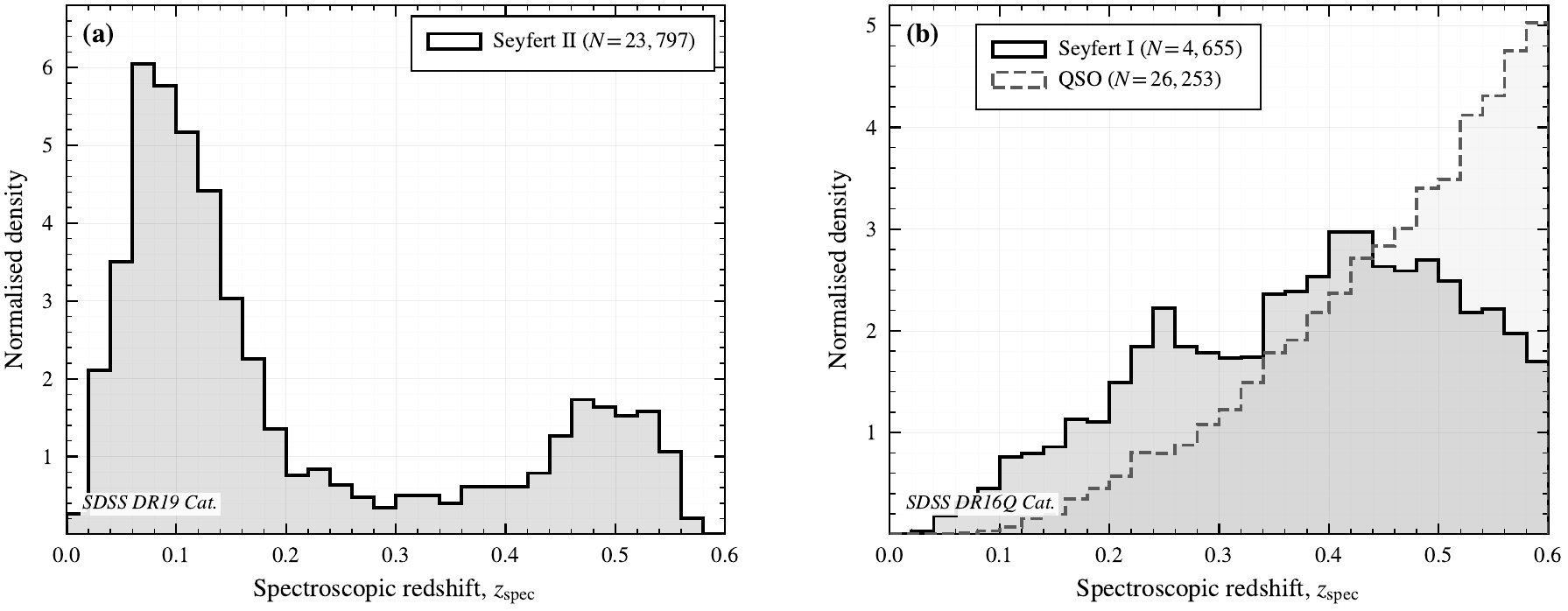}
    \caption{Spectroscopic redshift distributions of the final samples.
    Panel (a) shows the Seyfert II galaxies selected from the SDSS DR19
    catalogue, while panel (b) compares the Seyfert I and QSO samples
    selected from the SDSS DR16Q catalogue. The histograms are normalised
    to unit area. The final samples contain 23,797 Seyfert II galaxies,
    4,655 Seyfert I candidates, and 26,253 QSOs.}
    \label{fig:redshift_distributions}
\end{figure*}

Following the emission-line quality criteria of \citet{WuShen2022}, a Balmer line measurement was treated as usable when the broad component flux, flux uncertainty, logarithmic line luminosity, FWHM, and catalogue fit-pixel statistic were finite; the flux uncertainty was positive; the line signal-to-noise ratio exceeded 2; and the line luminosity satisfied \(38<\log_{10}(L_{\mathrm{line}}/\mathrm{erg\,s^{-1}})<48\). We additionally required more than half of the maximum number of pixels available for the corresponding line complex to be present. The maximum numbers are 264 pixels for the H$\alpha$ complex and 411 pixels for H$\beta$ in the Wu and Shen catalogue, so this requirement corresponds to more than 132 and 205.5 fitted pixels, respectively. A source was retained as a broad-line Type~I AGN when at least one usable Balmer line additionally satisfied

\begin{equation}
\mathrm{FWHM} > 1000~\mathrm{km\,s^{-1}}.
\label{eq:broad_line_selection}
\end{equation}

Objects assigned a visual classification other than QSO in DR16Q were excluded. The Balmer line and visual classification requirements resulted in an initial sample of 31,090 broad-line Type~I AGN.

The DR16Q catalogue redshift \(z_{\mathrm{DR16Q}}\) was then compared with the systemic redshift \(z_{\mathrm{SYS}}\) from \citet{WuShen2022}. We defined
\[
\Delta v =
c\,\frac{z_{\mathrm{DR16Q}}-z_{\mathrm{SYS}}}{1+z_{\mathrm{SYS}}},
\]
and required a finite systemic redshift within \(0<z_{\mathrm{SYS}}\leq0.6\), together with \(|\Delta v|\leq10{,}000~\mathrm{km\,s^{-1}}\). This deliberately permissive threshold was used only to remove large discrepancies between the catalogue and systemic redshifts; it was not intended as a physical or kinematic selection on the broad-line AGN population. This step removed sources without a valid systemic redshift comparison or with strongly discrepant redshift measurements, leaving 30,995 objects.

For the luminosity division, we used the DR16Q \texttt{M\_I} catalogue quantity, the extinction corrected absolute \(i\)-band magnitude normalised to \(z=2\) \citep{Lyke2020}. Sources with a valid value were divided at

\begin{equation}
M_i = -22.
\label{eq:mi_division}
\end{equation}

Objects with \(M_i > -22\) were assigned to the low luminosity broad-line Seyfert~I candidate sample, whereas objects with \(M_i \leq -22\) were assigned to the luminous QSO comparison sample. The \(M_i=-22\) boundary is used here only as an operational luminosity division within the spectroscopically selected broad-line population and does not replace the Balmer line selection. Before the final multiband completeness requirement, the two samples contained 4,660 Seyfert~I candidates and 26,276 QSOs.

The SDSS optical samples were subsequently matched to WISE, as described in Section~\ref{sec:wise_crossmatch}. Requiring valid SDSS \(ugriz\) and WISE \(W1\)--\(W4\) photometry resulted in final samples of 23,797 Seyfert~II galaxies, 4,655 Seyfert~I candidates, and 26,253 QSOs. Their spectroscopic redshift distributions are shown in Figure~\ref{fig:redshift_distributions}.

\subsection{Infrared Data and WISE Cross matching}
\label{sec:wise_crossmatch}

Mid-infrared photometry was obtained from the Wide-field Infrared Survey Explorer (WISE) all sky survey \citep{Wright2010}. WISE provides photometric measurements in four bands, \(W1\), \(W2\), \(W3\), and \(W4\), centred at approximately 3.4, 4.6, 12, and \(22~\mu\mathrm{m}\), respectively. These measurements extend the SDSS optical coverage into the mid-infrared and provide information on the stellar continuum and dust related emission of the selected AGN populations.

The SDSS positions were matched with the AllWISE Source Catalogue using a positional search radius of \(3~\mathrm{arcsec}\). This radius was adopted to account for the astrometric differences between the two surveys while limiting the number of unrelated associations \citep{Wright2010,Lang2016}. The cross matching was performed after the spectroscopic sample selection described in Section~\ref{sec:optical_sample_selection}. The definitions of the Seyfert~II, Seyfert~I, and QSO samples therefore depend on their SDSS spectroscopic properties rather than on the presence of a WISE counterpart.

For each accepted association, the catalogue measurements in the four WISE bands were combined with the SDSS \(ugriz\) photometry. Sources were required to have finite measurements in all nine bands. No photometric values were imputed. Objects with missing or nonfinite measurements in any of the adopted bands were excluded before the construction of the machine learning features.

The multiband completeness requirement retained all 23,797 objects in the Seyfert~II sample. In the samples derived from DR16Q, the number of Seyfert~I candidates decreased from 4,660 to 4,655, while the QSO sample decreased from 26,276 to 26,253. These catalogues constitute the final samples used in the subsequent analysis.

The WISE magnitudes were used without applying an additional correction for Galactic extinction. Interstellar attenuation is substantially lower at mid-infrared wavelengths than in the optical bands and is not expected to produce a significant change in the adopted WISE measurements \citep{Indebetouw2005}. The conversion of the WISE catalogue magnitudes to the photometric system used for feature construction is described in Section~\ref{sec:feature_construction}.

\begin{figure}[!htbp]
    \centering
    \includegraphics[
        width=\textwidth,
        trim=1.2cm 1.0cm 1.2cm 1.0cm,
        clip
    ]{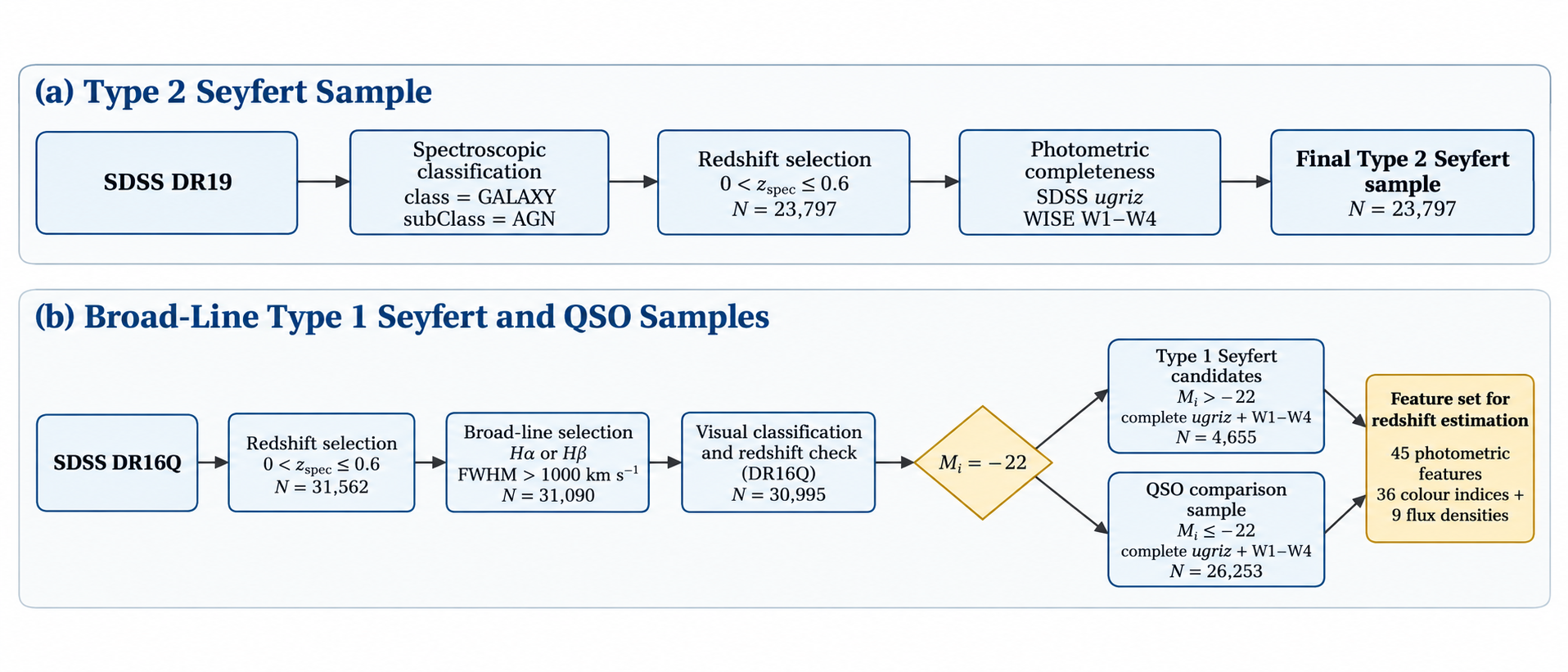}

    \caption{
    Overview of the sample selection and photometric preparation
    procedure. Panel (a) summarises the construction of the
    spectroscopically selected Seyfert~II sample from SDSS DR19.
    Panel (b) shows the selection of the broad-line Type~I AGN
    sample from SDSS DR16Q and its division into Seyfert~I
    candidates and the QSO comparison sample using the operational
    luminosity boundary \(M_i=-22\). The final catalogues contain
    complete SDSS \(ugriz\) and WISE \(W1\)--\(W4\) photometry.
    The same set of 45 photometric features, comprising 36 pairwise
    colours and nine flux densities, was constructed for all three
    samples.
    }
    \label{fig:sample_selection}
\end{figure}

\subsection{Photometric Feature Construction and Final Sample Definition}
\label{sec:feature_construction}

The same photometric representation was constructed for the Seyfert~II, Seyfert~I, and QSO samples. The input photometry consists of the five SDSS bands, \(u\), \(g\), \(r\), \(i\), and \(z\), together with the four WISE bands, \(W1\), \(W2\), \(W3\), and \(W4\). This provides nine broadband measurements spanning the optical and mid-infrared wavelength ranges.

Before constructing the features, all magnitudes were placed on a common AB magnitude system. The small SDSS offsets adopted in the analysis follow the standard SDSS photometric calibration convention \citep{York2000,Aihara2011}:

\begin{equation}
u_{\mathrm{AB}} = u_{\mathrm{SDSS}} - 0.040,
\qquad
z_{\mathrm{AB}} = z_{\mathrm{SDSS}} + 0.020,
\label{eq:sdss_ab_offsets}
\end{equation}

while the \(g\), \(r\), and \(i\) magnitudes were used without an additional offset. The WISE catalogue magnitudes, originally reported in the Vega system, were converted with the standard WISE Vega-to-AB offsets \citep{Jarrett2011}:

\begin{align}
W1_{\mathrm{AB}} &= W1_{\mathrm{Vega}} + 2.699, \nonumber\\
W2_{\mathrm{AB}} &= W2_{\mathrm{Vega}} + 3.339, \nonumber\\
W3_{\mathrm{AB}} &= W3_{\mathrm{Vega}} + 5.174, \nonumber\\
W4_{\mathrm{AB}} &= W4_{\mathrm{Vega}} + 6.620.
\label{eq:wise_ab_offsets}
\end{align}

A total of 36 colour indices was then generated by taking every unique pairwise difference among the nine AB magnitudes. For two bands \(i\) and \(j\), each colour was defined as

\begin{equation}
C_{ij} = m_i - m_j,
\qquad i < j.
\label{eq:pairwise_colours}
\end{equation}

The number of unique pairwise band combinations is therefore

\begin{equation}
\binom{9}{2} = 36.
\end{equation}

The use of all pairwise colours extends the input representation beyond adjacent band colours and provides a uniform description of the broadband spectral shape across the optical and mid-infrared bands. Because these colours are algebraically redundant, they are used here as a predictive representation for the tree-based models rather than as independent physical observables.

The nine AB magnitudes were also converted to flux densities. For each band, the flux density in mJy was calculated from

\begin{equation}
f_{\nu}\,(\mathrm{mJy})
=
3.631 \times 10^{6}
\,10^{-0.4m_{\mathrm{AB}}}.
\label{eq:ab_flux_conversion}
\end{equation}

The flux density terms retain information on the overall brightness of each source, whereas the colour indices primarily describe the relative shape of the observed spectral energy distribution. Combining the 36 colours with the nine flux densities produced a 45 feature photometric vector for each object:

\begin{equation}
N_{\mathrm{features}}
=
36\ \mathrm{colours}
+
9\ \mathrm{fluxes}
=
45.
\end{equation}

The same sequence of conversions and feature definitions was applied to all three source samples. The spectroscopic redshift was retained as the target variable for the regression analysis but was not included among the input features. The QSO sample was processed using the same photometric representation so that it could be used as an external evaluation sample for the Seyfert subtype classifier.

After requiring finite values for all 45 features, the final analysis catalogues contained 23,797 Seyfert~II galaxies, 4,655 Seyfert~I candidates, and 26,253 QSOs. The corresponding feature matrices have dimensions \(23{,}797 \times 45\), \(4{,}655 \times 45\), and \(26{,}253 \times 45\), respectively.

\section{Machine Learning Methods}
\label{sec:methods}

\subsection{Random Forest Regression}
\label{sec:random_forest}

Random Forest regression combines predictions from many decision trees fitted to bootstrap samples while considering a random subset of features at each split \citep{Breiman2001}. The method can represent nonlinear relations and interactions without requiring a prescribed functional form and has been used extensively for photometric redshift estimation \citep{Carliles2010,CarrascoKind2013}. In this study, Random Forest regression was used for the primary Seyfert~II photo-\(z\) model and for the controlled transfer experiments between Seyfert subtypes. The same Random Forest configuration was used in both transfer directions so that differences in performance were not driven by the choice of regression algorithm. The adopted hyperparameters and validation procedure are described in Section~\ref{sec:model_training}.

\subsection{XGBoost Regression}
\label{sec:xgboost}

XGBoost is a gradient boosted decision tree method in which trees are added sequentially to reduce the residual errors of the preceding ensemble while regularisation controls model complexity \citep{Chen2016}. This combination is well suited to nonlinear relations among correlated photometric features. XGBoost was selected as the primary regressor for the Seyfert~I sample after model comparison, and its reported performance was obtained from the outer folds of the nested cross validation procedure described in Section~\ref{sec:model_training}. Transferability between the two Seyfert subtypes was evaluated separately with identically configured Random Forest regressors in both directions.

\subsection{Performance Metrics}
\label{sec:performance_metrics}

Photometric redshift performance was evaluated with the coefficient of determination \(R^2\), mean absolute error (MAE), root mean square error (RMSE), bias, normalised median absolute deviation, and the outlier fraction. For each source, the normalised residual was defined as
\begin{equation}
\delta_i =
\frac{
z_{\mathrm{phot},i}-z_{\mathrm{spec},i}
}{
1+z_{\mathrm{spec},i}
}.
\label{eq:normalised_residual}
\end{equation}

The bias is the mean of \(\delta_i\), with positive values indicating an average overestimate of redshift and negative values indicating an underestimate. The residual scatter was quantified by
\begin{equation}
\sigma_{\mathrm{NMAD}}
=
1.4826\,
\mathrm{median}
\left(
\left|
\delta_i-\mathrm{median}(\delta)
\right|
\right),
\label{eq:nmad}
\end{equation}
which is less sensitive to a small number of large residuals than a standard deviation and is widely used in photo-\(z\) studies \citep{Salvato2019}. A source was classified as an outlier when
\begin{equation}
|\delta_i|>0.15.
\label{eq:outlier_definition}
\end{equation}
The outlier fraction \(\eta\) is reported as the percentage of evaluated sources satisfying this criterion. The same metric definitions were used for within-subtype validation, cross-subtype transfer, matched controls, feature ablations, and the independent classifier gated regression test.

\subsection{Model Training, Cross Subtype Evaluation, and Selective Prediction}
\label{sec:model_training}

Within-subtype performance was estimated from out-of-fold predictions. The Seyfert~II Random Forest model used five-fold cross validation. For the Seyfert~I XGBoost model, five outer folds formed a nested cross validation procedure so that hyperparameter selection was performed only within each outer training sample. The combined outer fold predictions therefore provided one prediction for every source without using that source for either model fitting or, for XGBoost, hyperparameter selection.

To preserve the approximate spectroscopic redshift distribution across folds, the Seyfert~II sample was divided into 20 redshift quantile groups and the Seyfert~I sample into 10 groups before fold assignment. Stratified folds were shuffled with random seed 42. Spectroscopic redshift was used only for stratification and as the regression target, never as an input feature.

All Random Forest regressors used 500 trees, \texttt{max\_features="sqrt"}, random seed 42, and otherwise default scikit-learn parameters. This configuration was used for the Seyfert~II model, the Seyfert~I Random Forest reference model, and both directions of the controlled transfer experiment.

For the Seyfert~I XGBoost analysis, hyperparameters were selected with randomised optimisation using MAE as the objective. Within each outer training fold, an inner three-fold cross validation search evaluated 15 randomly sampled configurations. The search sampled
\(n_{\mathrm{estimators}}\in\{500,750,1000,1250,1500\}\),
learning rates \(\{0.01,0.02,0.03,0.05,0.08\}\),
maximum depths \(\{3,4,5,6\}\),
minimum child weights \(\{1,2,3,5,8\}\),
row and feature subsampling fractions \(\{0.70,0.80,0.90,1.00\}\),
L1 regularisation values \(\{0,0.01,0.05,0.10,0.50\}\),
and L2 regularisation values \(\{0.5,1.0,2.0,5.0\}\).

A separate five-fold randomised search on the complete Seyfert~I sample, evaluating 50 configurations from the same search space, was used only to define a reproducible final model configuration for subsequent fitting. The selected model used 500 trees, learning rate 0.02, maximum depth 6, minimum child weight 8, row subsampling 0.80, feature subsampling 0.90, L1 regularisation 0.05, and L2 regularisation 5.0, with the squared error objective and random seed 42. This final configuration was not used to estimate the reported cross validated performance. The Seyfert~I metrics reported in this work come exclusively from the outer fold predictions of nested cross validation.

To test transferability between subtypes, identically configured Random Forest models were used in both directions. One model was fitted to the complete Seyfert~I sample and evaluated independently on Seyfert~II, and the reverse experiment was then performed by fitting to the complete Seyfert~II sample and evaluating on Seyfert~I. Both experiments used the same 45 photometric features.

\begin{table*}[!t]
\centering
\caption{Photometric redshift performance of the regression models
developed separately for each Seyfert subtype. The Seyfert~II values
come from out-of-fold predictions from five-folds, whereas the Seyfert~I
values come from the outer fold predictions of nested cross validation.
The outlier percentage corresponds to sources satisfying
\(\left|z_{\mathrm{phot}}-z_{\mathrm{spec}}\right|/
(1+z_{\mathrm{spec}})>0.15\).}
\label{tab:subtype_photoz_performance}
\small
\begin{tabular}{llrrrrrrr}
\hline
Sample &
Regressor &
\(N\) &
\(R^2\) &
MAE &
RMSE &
Bias &
\(\sigma_{\mathrm{NMAD}}\) &
Outlier (\%) \\
\hline
Seyfert~I  &
XGBoost &
4,655 &
0.82662 &
0.03852 &
0.05532 &
0.00169 &
0.02962 &
0.988 \\

Seyfert~II &
Random Forest &
23,797 &
0.96538 &
0.02008 &
0.03060 &
0.00058 &
0.01691 &
0.206 \\
\hline
\end{tabular}
\end{table*}

A matched control was used to test whether the transfer behaviour was driven mainly by differences in sample size or redshift distribution. The interval \(0<z_{\mathrm{spec}}\leq0.6\) was divided into bins of width \(\Delta z=0.05\), and equal numbers of Seyfert~I and Seyfert~II sources were sampled within each bin. Bins with fewer than ten sources from either subtype were excluded, yielding 4,521 sources per subtype. The matching was repeated independently 50 times. Within-subtype performance was evaluated with five-fold out-of-fold predictions stratified by redshift bin, while cross-subtype performance was measured by fitting on the complete matched sample of one subtype and evaluating on the matched sample of the other. Results are reported as the median and the 2.5th to 97.5th percentile range across the 50 realisations.

A Random Forest classifier was then trained as a selection step before the Seyfert~II regressor. The classification sample was matched between Seyfert~I and Seyfert~II in the same \(\Delta z=0.05\) bins and contained 4,521 sources from each subtype. An 80:20 split, stratified jointly by source class and redshift bin, produced a development set containing 3,617 Seyfert~I and 3,616 Seyfert~II sources and an untouched final test set containing 904 Seyfert~I and 905 Seyfert~II sources. The classifier used 500 trees, \texttt{max\_features="sqrt"}, balanced class weights, and random seed 42.

Five fold out-of-fold probabilities were generated within the development set. Thresholds from 0.500 to 0.990 in steps of 0.001 were examined. Values with Seyfert~II precision below 0.95 were discarded, and the remaining threshold with the highest Seyfert~II recall was selected, with ties resolved by higher \(F_1\) and then the lower threshold. This procedure gave
\begin{equation}
P(\mathrm{Sy\,II}) \geq 0.620.
\label{eq:classifier_threshold}
\end{equation}
The threshold was fixed before the classifier was fitted to the complete development set and evaluated on the untouched Seyfert test sample. The same fitted classifier and threshold were then applied to the 26,253 QSO comparison sources, which were excluded from matching, classifier training, and threshold selection.

Two feature ablation tests were performed with the controlled Random Forest transfer setup. The first compared the 36 pairwise colours, the nine broadband flux densities, and the full 45 feature representation. The second compared SDSS only features, WISE only features, and combined SDSS+WISE features. The training samples, Random Forest configuration, and evaluation procedure were held fixed so that changes in performance reflected the adopted photometric inputs.

Finally, the classifier gate was tested directly against independent photo-\(z\) accuracy. The 905 Seyfert~II objects in the untouched classifier test sample were removed from the regression training data. A Random Forest regressor with the same configuration as the transfer experiment was fitted to the remaining 22,892 Seyfert~II sources and applied to the complete classifier test sample. Regression performance was then evaluated separately for sources accepted and rejected by the fixed classifier threshold. None of the Seyfert~II sources used in this gate evaluation was included in the regression training sample. As an additional test of redshift-domain dependence, we evaluated the behaviour of the subtype-specific regressors beyond the redshift range represented during training. For each subtype, the model was trained only on sources with \(0<z_{\mathrm{spec}}\leq0.5\) and was then evaluated on sources with \(0.5<z_{\mathrm{spec}}\leq0.6\). This produced 3,672 training and 983 test sources for Seyfert~I, and 21,715 training and 2,082 test sources for Seyfert~II. The same XGBoost configuration adopted for Seyfert~I and the same Random Forest configuration adopted for Seyfert~II were retained. Because the evaluation interval spans a narrow range in spectroscopic redshift, the interpretation of this stress test focused primarily on MAE, bias, \(\sigma_{\mathrm{NMAD}}\), and the outlier fraction rather than on \(R^2\).

\section{Results}
\label{sec:results}

\subsection{Photometric Redshift Performance within Each Seyfert Subtype}
\label{sec:subtype_photoz_results}

The photometric redshift models developed for each Seyfert subtype
were first evaluated within the samples from which they were
constructed. For the Seyfert~II Random Forest model, five-fold cross
validation was used to obtain one out-of-fold prediction for every
source. For the Seyfert~I XGBoost model, performance was evaluated
from the outer fold predictions of the nested cross validation
procedure described in Section~\ref{sec:model_training}. In both
cases, each reported prediction was produced without using the
corresponding source during model fitting and, for the Seyfert~I
analysis, also without using it during hyperparameter selection.
The resulting performance statistics are presented in
Table~\ref{tab:subtype_photoz_performance}.

For the Seyfert~I sample, the XGBoost regressor achieved
\(R^2=0.82662\), with an MAE of \(0.03852\) and an RMSE of
\(0.05532\). The mean normalised residual remained close to zero,
with a bias of \(0.00169\), while the residual scatter was
\(\sigma_{\mathrm{NMAD}}=0.02962\). The fraction of sources satisfying
the adopted outlier criterion was \(0.988\%\).

The Random Forest model produced smaller prediction errors within the
Seyfert~II sample. It yielded \(R^2=0.96538\), an MAE of \(0.02008\),
and an RMSE of \(0.03060\). The bias was \(0.00058\), and the residual
scatter was \(\sigma_{\mathrm{NMAD}}=0.01691\). The outlier fraction was
\(0.206\%\). 

The lower errors obtained for the Seyfert~II sample provide a descriptive
comparison between the two models developed separately for the subtypes. However, the two
results were obtained from samples with different sizes and redshift
distributions and with different selected regression algorithms.
Consequently, this comparison alone does not isolate the effect of
Seyfert subtype. The dependence of model performance on the evaluation
subtype is examined separately in Section~\ref{sec:cross_subtype_results}.

\begin{table}[!htbp]
\centering
\caption{Controlled Random Forest evaluation within and between the Seyfert subtypes. The results within each subtype were obtained from out-of-fold predictions from five-folds. For the independent evaluations, the model was fitted to the complete training subtype and evaluated on the other subtype. Identical Random Forest hyperparameters and the same 45 photometric features were used in all four experiments.}
\label{tab:cross_subtype}

\scriptsize
\setlength{\tabcolsep}{2pt}
\renewcommand{\arraystretch}{1.12}

\begin{tabular}{@{}lllccccccc@{}}
\toprule
\shortstack{Training\\sample} &
\shortstack{Evaluation\\sample} &
Evaluation &
$N_{\rm eval}$ &
$R^2$ &
MAE &
RMSE &
Bias &
$\sigma_{\rm NMAD}$ &
\shortstack{Outlier\\(\%)} \\
\midrule

Seyfert~I &
Seyfert~I &
OOF CV &
4,655 &
0.81996 &
0.03962 &
0.05638 &
0.00148 &
0.03064 &
1.010 \\

Seyfert~I &
Seyfert~II &
Independent &
23,797 &
0.84638 &
0.04717 &
0.06445 &
$-0.00348$ &
0.04155 &
0.513 \\

Seyfert~II &
Seyfert~II &
OOF CV &
23,797 &
0.96538 &
0.02008 &
0.03060 &
0.00058 &
0.01691 &
0.206 \\

Seyfert~II &
Seyfert~I &
Independent &
4,655 &
0.55915 &
0.06819 &
0.08822 &
0.02167 &
0.05336 &
3.996 \\

\bottomrule
\end{tabular}

\end{table}

\subsection{Evaluation Between Seyfert Subtypes}
\label{sec:cross_subtype_results}

A controlled cross-subtype evaluation was performed to separate the
effect of Seyfert subtype from differences caused by the choice of
regression algorithm. Random Forest models with identical
hyperparameters and the same set of 45 photometric features were
therefore used in both evaluation directions. The within-subtype
five-fold out-of-fold results provide the reference performance,
while the transfer results were obtained by fitting the model
to the complete source subtype and evaluating it independently on
the other subtype. The results are listed in
Table~\ref{tab:cross_subtype}.

Within the Seyfert~I sample, the Random Forest model achieved
\(R^2=0.81996\), an MAE of \(0.03962\), an RMSE of \(0.05638\),
and \(\sigma_{\mathrm{NMAD}}=0.03064\). When the model was fitted
to the complete Seyfert~I sample and evaluated on the Seyfert~II
sample, the MAE increased to \(0.04717\), the RMSE to \(0.06445\),
and the NMAD to \(0.04155\). The bias remained small and negative,
at \(-0.00348\), while the outlier fraction was \(0.513\%\).

The Random Forest model evaluated within the Seyfert~II sample
produced \(R^2=0.96538\), an MAE of \(0.02008\), an RMSE of
\(0.03060\), and \(\sigma_{\mathrm{NMAD}}=0.01691\). Its outlier
fraction was \(0.206\%\). In contrast, when the model fitted to the
complete Seyfert~II sample was evaluated on the Seyfert~I sample,
the performance decreased to \(R^2=0.55915\). The MAE and RMSE
increased to \(0.06819\) and \(0.08822\), respectively, while the
NMAD increased to \(0.05336\). A positive bias of \(0.02167\) was
measured, and the outlier fraction increased to \(3.996\%\).

The transfer behaviour between the Seyfert subtypes was therefore asymmetric. The model
trained on Seyfert~II sources showed a substantially larger loss of
accuracy when evaluated on Seyfert~I sources than the model trained
on Seyfert~I sources showed when evaluated on Seyfert~II sources. The apparently higher \(R^2\) obtained for the model trained on
Seyfert~I sources and evaluated on the Seyfert~II sample should not be
interpreted as an improvement in absolute predictive accuracy,
because the MAE, RMSE, and NMAD all increased. Moreover,
\(R^2\) depends on the variance of the spectroscopic redshift
distribution in the evaluated sample and is therefore not directly
comparable between samples with different redshift distributions.

\begin{figure*}[!t]
\centering
\includegraphics[width=\textwidth]
{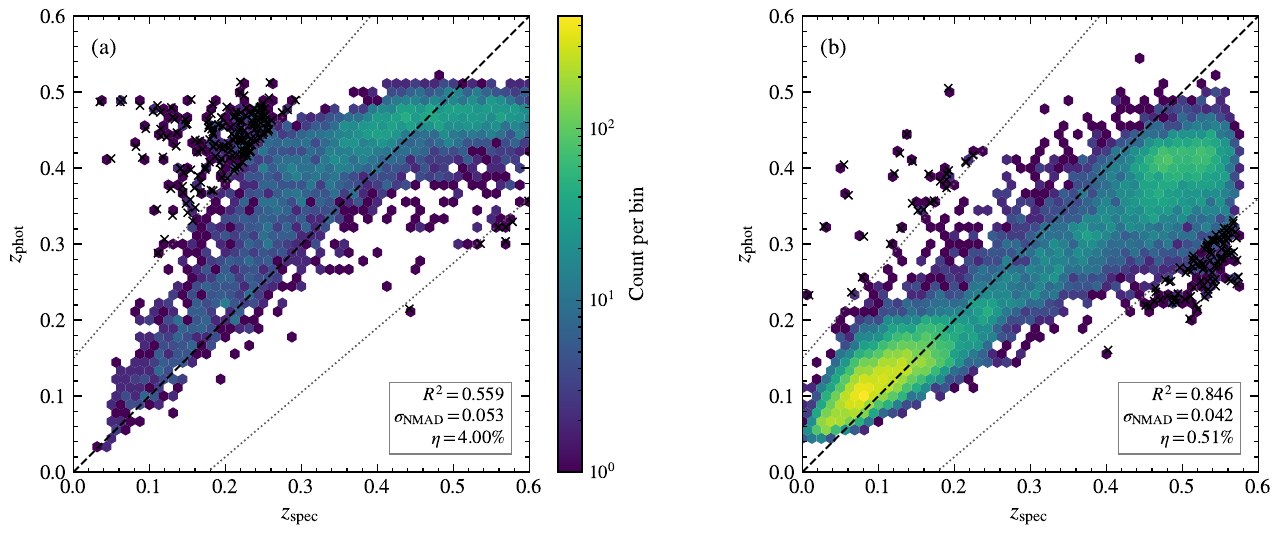}
\caption{Controlled cross-subtype photometric redshift evaluation
using Random Forest models with identical hyperparameters and the
same 45 photometric features. Panel (a) shows the model fitted to
the Seyfert~II sample and evaluated independently on the
Seyfert~I sample. Panel (b) shows the model fitted to the
Seyfert~I sample and evaluated independently on the
Seyfert~II sample. The hexagonal bins indicate the number of
sources on a logarithmic scale. The dashed line represents
\(z_{\mathrm{phot}}=z_{\mathrm{spec}}\), while the dotted lines
mark the adopted outlier boundary,
\(\lvert z_{\mathrm{phot}}-z_{\mathrm{spec}}\rvert/
(1+z_{\mathrm{spec}})=0.15\). Crosses identify sources satisfying
the outlier criterion.}
\label{fig:cross_subtype_rf}
\end{figure*}

The corresponding predicted and spectroscopic redshifts are shown
in Figure~\ref{fig:cross_subtype_rf}. Although the controlled
experiment removes algorithm choice as a source of the observed
difference, the two samples also have different redshift
distributions. The observed asymmetry therefore cannot be
attributed exclusively to the physical distinction between the
two Seyfert subtypes.

\begin{table}[!htbp]
\centering
\caption{Random Forest performance across 50 independently constructed realisations matched in redshift and sample size. Each realisation contained 4,521 Seyfert I and 4,521 Seyfert II sources. Values are reported as the median, with the 2.5th--97.5th percentile range given in square brackets. Within-subtype results were obtained from out-of-fold predictions from five-folds, whereas transfer results were obtained by training on one matched subtype and evaluating independently on the other.}
\label{tab:matched_cross_subtype}

\scriptsize
\setlength{\tabcolsep}{2pt}
\renewcommand{\arraystretch}{1.15}

\begin{tabular}{@{}llcccc@{}}
\toprule
\shortstack{Training\\sample} &
\shortstack{Evaluation\\sample} &
MAE &
Bias &
$\sigma_{\rm NMAD}$ &
\shortstack{Outlier\\(\%)} \\
\midrule

Seyfert~I &
Seyfert~I &
\shortstack{$0.03925$\\$[0.03904,\,0.03955]$} &
\shortstack{$0.00141$\\$[0.00128,\,0.00152]$} &
\shortstack{$0.03026$\\$[0.02985,\,0.03067]$} &
\shortstack{$0.973$\\$[0.907,\,1.079]$} \\

Seyfert~I &
Seyfert~II &
\shortstack{$0.07391$\\$[0.07265,\,0.07613]$} &
\shortstack{$-0.04464$\\$[-0.04586,\,-0.04367]$} &
\shortstack{$0.04136$\\$[0.04034,\,0.04277]$} &
\shortstack{$1.239$\\$[1.055,\,1.416]$} \\

Seyfert~II &
Seyfert~II &
\shortstack{$0.03470$\\$[0.03420,\,0.03515]$} &
\shortstack{$0.00119$\\$[0.00103,\,0.00133]$} &
\shortstack{$0.02907$\\$[0.02855,\,0.02992]$} &
\shortstack{$0.243$\\$[0.155,\,0.393]$} \\

Seyfert~II &
Seyfert~I &
\shortstack{$0.06896$\\$[0.06600,\,0.07133]$} &
\shortstack{$0.02518$\\$[0.02224,\,0.02852]$} &
\shortstack{$0.05833$\\$[0.05527,\,0.06145]$} &
\shortstack{$3.893$\\$[3.416,\,4.497]$} \\

\bottomrule
\end{tabular}

\end{table}

\subsection{Evaluation with Matched Redshift Distributions and Sample Sizes}
\label{sec:matched_transfer}

To control for differences in sample size and spectroscopic redshift
distribution, the transfer analysis was repeated using
50 independently constructed matched realisations. Each
realisation contained 4,521 Seyfert~I and 4,521 Seyfert~II
sources, with equal numbers selected from the two subtypes within
spectroscopic redshift bins of width $\Delta z=0.05$. The median results and the corresponding
$2.5$th--$97.5$th percentile ranges are presented in
Table~\ref{tab:matched_cross_subtype}.

Within the matched samples, the Random Forest trained on Seyfert~I
sources achieved a median MAE within the same subtype of $0.03925$ and
$\sigma_{\mathrm{NMAD}}=0.03026$. When evaluated on the matched
Seyfert~II sample, its median MAE increased to $0.07391$ and its
bias shifted from $0.00141$ to $-0.04464$. The corresponding
median $\sigma_{\mathrm{NMAD}}$ was $0.04136$, with an outlier
fraction of $1.239\%$.

The model trained on Seyfert~II sources achieved a median MAE within the same subtype
of $0.03470$, $\sigma_{\mathrm{NMAD}}=0.02907$, and an outlier
fraction of $0.243\%$. When transferred to the matched
Seyfert~I sample, the median MAE increased to $0.06896$,
$\sigma_{\mathrm{NMAD}}$ increased to $0.05833$, and the outlier
fraction increased to $3.893\%$.

The loss of performance after transfer between the Seyfert subtypes therefore persisted after
matching the two samples in both size and spectroscopic redshift
distribution. The two transfer directions nevertheless exhibited
different error structures. The model trained on Seyfert~I sources produced
a stronger systematic underestimation for Seyfert~II sources,
whereas the model trained on Seyfert~II sources produced a broader residual
distribution and a substantially larger outlier
fraction for Seyfert~I sources.

\subsection{Robustness to the Adopted Photometric Features}
\label{sec:feature_ablation_results}

The controlled transfer experiment was repeated after separating
the photometric representation into different feature families and
wavelength ranges. The results are summarised in
Table~\ref{tab:feature_ablation_transfer}. When only the 36
pairwise colours were used, the transfer from Seyfert~I to
Seyfert~II yielded an MAE of \(0.05107\) and
\(\sigma_{\mathrm{NMAD}}=0.04685\), while the reverse transfer
yielded an MAE of \(0.07618\),
\(\sigma_{\mathrm{NMAD}}=0.06253\), and an outlier fraction of
\(5.843\%\). The transfer loss therefore remained present after
the broadband flux densities were removed.

\begin{table*}[!t]
\centering
\caption{Cross-subtype Random Forest performance for the photometric
feature ablation tests. The regression algorithm, training samples,
and evaluation procedure were kept fixed while the adopted input
features were changed.}
\label{tab:feature_ablation_transfer}
\small
\setlength{\tabcolsep}{4pt}
\renewcommand{\arraystretch}{1.15}
\begin{tabular}{lrrrrrrr}
\hline
Feature set &
\(N_{\mathrm{feat}}\) &
\multicolumn{3}{c}{Seyfert~I \(\rightarrow\) Seyfert~II} &
\multicolumn{3}{c}{Seyfert~II \(\rightarrow\) Seyfert~I} \\
\cmidrule(lr){3-5}\cmidrule(lr){6-8}
&
&
MAE &
\(\sigma_{\mathrm{NMAD}}\) &
Outlier (\%) &
MAE &
\(\sigma_{\mathrm{NMAD}}\) &
Outlier (\%) \\
\hline
Colours only      & 36 & 0.05107 & 0.04685 & 1.408 & 0.07618 & 0.06253 & 5.843 \\
Flux densities only & 9 & 0.04856 & 0.04301 & 0.626 & 0.06837 & 0.05482 & 3.845 \\
SDSS only         & 15 & 0.04812 & 0.04136 & 0.651 & 0.06644 & 0.05060 & 4.039 \\
WISE only         & 10 & 0.08696 & 0.07814 & 11.485 & 0.09386 & 0.08729 & 8.657 \\
SDSS+WISE         & 45 & 0.04717 & 0.04155 & 0.513 & 0.06819 & 0.05336 & 3.996 \\
\hline
\end{tabular}
\end{table*}

Using the nine flux densities alone also preserved the asymmetric
transfer behaviour. The Seyfert~I to Seyfert~II transfer produced
an MAE of \(0.04856\) and
\(\sigma_{\mathrm{NMAD}}=0.04301\), whereas the Seyfert~II to
Seyfert~I transfer produced an MAE of \(0.06837\) and
\(\sigma_{\mathrm{NMAD}}=0.05482\). The complete 45 feature
representation provided the lowest MAE in the Seyfert~I to
Seyfert~II direction and retained the main transfer behaviour in
the reverse direction. These comparisons show that the observed
cross-subtype degradation is not caused solely by the inclusion
of apparent flux density information.

A complementary wavelength test showed that the SDSS optical
photometry contained most of the predictive information relevant
to the present redshift range. With SDSS features alone, the MAE
was \(0.04812\) for Seyfert~I to Seyfert~II transfer and
\(0.06644\) for Seyfert~II to Seyfert~I transfer. The corresponding
WISE only values were \(0.08696\) and \(0.09386\), respectively,
with substantially larger residual scatter and outlier fractions.
Importantly, the transfer degradation remained pronounced when
only the SDSS bands were used. It therefore cannot be attributed
solely to the inclusion of mid-infrared photometry.

\subsection{Selective Prediction with a Seyfert Subtype Classifier}
\label{sec:selective_prediction_results}

The Seyfert subtype classifier was used as a preliminary selection
step before applying the Seyfert~II photometric redshift model. A
source was accepted for Seyfert~II photo-\(z\) estimation when its
predicted Seyfert~II probability satisfied
\(P(\mathrm{Sy\,II}) \geq 0.620\). The classifier was trained to
distinguish Seyfert~I and Seyfert~II sources; the QSO sample was not
included in the classifier training and was used only as an external
evaluation sample.

At the adopted probability threshold, the out-of-fold classification
results yielded a Seyfert~II precision of \(0.95031\), a Seyfert~II
recall of \(0.88855\), and a Seyfert~I specificity of \(0.95355\).
These values indicate that the selected sample retained most of the
Seyfert~II sources while admitting only a small fraction of
Seyfert~I sources.

The same threshold was then applied to the independent
evaluation samples. Of the 905 Seyfert~II sources, 811 were
accepted, corresponding to an acceptance rate of \(89.61\%\).
By comparison, only 37 of the 904 Seyfert~I sources were
accepted, giving a false acceptance rate of \(4.09\%\).
When applied to the external QSO comparison sample, the
classifier accepted 651 of the 26,253 sources, corresponding
to \(2.48\%\). The complete acceptance and rejection statistics
are listed in Table~\ref{tab:classifier_performance}.

\begin{table}[!t]
\centering
\caption{Performance of the Seyfert subtype classifier at the
adopted threshold \(P(\mathrm{Sy\,II}) \geq 0.620\).
The out-of-fold statistics were obtained from the
classification sample containing Seyfert~I and Seyfert~II sources.
The QSO catalogue was not included in classifier training
and was used only as an external evaluation sample.}
\label{tab:classifier_performance}

\small
\renewcommand{\arraystretch}{1.15}

\begin{tabular}{lc}
\hline
\multicolumn{2}{c}{Out of fold classification performance} \\
\hline
Metric & Value \\
\hline
Seyfert~II precision   & 0.95031 \\
Seyfert~II recall      & 0.88855 \\
Seyfert~I specificity  & 0.95355 \\
\hline
\end{tabular}

\vspace{0.6em}

\begin{tabular}{lcc}
\hline
\multicolumn{3}{c}{Independent sample evaluation} \\
\hline
Sample & Accepted / total & Acceptance (\%) \\
\hline
Seyfert~II & \(811 / 905\)       & 89.61 \\
Seyfert~I  & \(37 / 904\)        & 4.09  \\
QSO        & \(651 / 26{,}253\)  & 2.48  \\
\hline
\end{tabular}

\end{table}

The low acceptance rates for the Seyfert~I and QSO samples show that
the classifier can reduce the application of the Seyfert~II
photometric redshift model to sources whose photometry is consistent
with the Seyfert~II training sample. The result should be interpreted
within the scope of the present analysis: the classifier distinguishes
Seyfert~I and Seyfert~II sources within a preselected low-redshift AGN
sample and was not designed as a general classifier for stars, inactive
galaxies, or unrestricted survey catalogues.

The selective classification step was also tested directly against
photometric redshift accuracy. For this experiment, the 905
Seyfert~II sources in the untouched classifier test sample were
excluded from the Seyfert~II regression training set. The regressor
was therefore trained on the remaining 22,892 Seyfert~II sources
before being applied to the independent test sample.

Among the 905 Seyfert~II sources, the classifier accepted 811 and
rejected 94. The accepted sources showed an MAE of \(0.03320\),
\(\sigma_{\mathrm{NMAD}}=0.02741\), and an outlier fraction of
\(0.247\%\). In contrast, the rejected sources had an MAE of
\(0.04986\), \(\sigma_{\mathrm{NMAD}}=0.04314\), and an outlier
fraction of \(1.064\%\). When all 905 Seyfert~II sources were
considered without selective filtering, the corresponding values
were an MAE of \(0.03493\),
\(\sigma_{\mathrm{NMAD}}=0.02881\), and an outlier fraction of
\(0.331\%\). The corresponding photometric redshift performance statistics for the full, accepted, and rejected subsets are presented in Table~\ref{tab:gate_photoz}. The classifier therefore preferentially retained
sources for which the Seyfert~II redshift relation was more
reliable.

\begin{table}[!t]
\centering
\caption{Photometric redshift performance of the Seyfert~II
regressor on the independent classifier test sample. The 905
Seyfert~II sources used for this evaluation were excluded from
regression training.}
\label{tab:gate_photoz}
\small
\renewcommand{\arraystretch}{1.15}
\begin{tabular}{lrrrrr}
\hline
Sample & \(N\) & \(R^2\) & MAE &
\(\sigma_{\mathrm{NMAD}}\) & Outlier (\%) \\
\hline
All      & 905 & 0.8722 & 0.03493 & 0.02881 & 0.331 \\
Accepted & 811 & 0.8885 & 0.03320 & 0.02741 & 0.247 \\
Rejected & 94  & 0.6759 & 0.04986 & 0.04314 & 1.064 \\
\hline
\end{tabular}
\end{table}

\FloatBarrier

\subsection{Redshift Extrapolation Test}
\label{sec:redshift_extrapolation}

When the training samples were restricted to
\(0<z_{\mathrm{spec}}\leq0.5\), both subtype-specific regressors
showed a clear loss of accuracy when evaluated above the training
redshift limit. For Seyfert~I, the 983 sources with
\(0.5<z_{\mathrm{spec}}\leq0.6\) yielded an MAE of \(0.10510\),
a bias of \(-0.06766\), \(\sigma_{\mathrm{NMAD}}=0.02610\), and
an outlier fraction of \(0.712\%\). For Seyfert~II, the 2,082
test sources yielded an MAE of \(0.07396\), a bias of
\(-0.04822\), \(\sigma_{\mathrm{NMAD}}=0.01602\), and an outlier
fraction of \(0.096\%\). The Seyfert~II test sample extended from
\(z_{\mathrm{spec}}=0.5000\) to \(0.5738\).

Despite the comparatively small NMAD and outlier fractions, the
substantially larger MAE values and negative biases show that both
models systematically underestimated redshift above the range
represented during training. The median predicted redshifts were
\(0.4511\) for Seyfert~I and \(0.4615\) for Seyfert~II, compared
with median spectroscopic redshifts of \(0.5465\) and \(0.5283\),
respectively.

\section{Discussion}
\label{sec:discussion}

The central result of this study is that a photometric redshift relation calibrated within one low-redshift Seyfert subtype cannot be assumed to retain the same accuracy when applied to the other. This conclusion is supported by the controlled transfer experiment, by 50 matched realisations that equalise sample size and spectroscopic redshift distribution, and by feature ablations that preserve the transfer loss after the input representation is restricted. The classifier analysis further shows that subtype probability can be used as a practical indicator of whether the specialised Seyfert~II regressor is being applied within a more reliable region of photometric parameter space.

\subsection{Photometric Redshift Performance within the Seyfert Subtypes}
\label{sec:discussion_subtype_performance}

The subtype specific models show a clear difference in within sample performance, with substantially lower residual scatter and prediction errors for the Seyfert~II sample than for the Seyfert~I sample. The Seyfert~I Random Forest reference model closely reproduced the XGBoost result, indicating that this contrast is not primarily a consequence of choosing different tree-based regressors.

The lower scatter in the Seyfert~II sample is consistent with the stronger contribution of host galaxy light expected when the nucleus is obscured. Host dominated optical SEDs can preserve a more regular relation between broadband shape and redshift, whereas broad-line Type I AGN contain a more variable mixture of host light, unobscured nuclear continuum, broad emission lines, and infrared emission. Previous AGN photo-\(z\) studies have similarly found that host-dominated and lower luminosity AGN can approach the performance obtained for inactive galaxies, while AGN-dominated sources tend to show broader and more degenerate colour redshift relations \citep{Salvato2009,Salvato2011,Brescia2021}.

Photometric variability may contribute to the broader Seyfert~I residual distribution because the SDSS and WISE measurements are not contemporaneous. The present analysis does not isolate variability from the other sources of photometric diversity, so it should be regarded as one possible contributor rather than a demonstrated cause. The unusually low Seyfert~II outlier fraction must also be interpreted within the restricted observational domain of this work. All samples are spectroscopically defined, limited to \(0<z_{\mathrm{spec}}\leq0.6\), and required to have finite measurements in all nine adopted bands. In particular, the complete SDSS+WISE requirement may favour objects with better measured mid-infrared photometry and does not reproduce the incompleteness expected in a general survey catalogue.

\subsection{Transferability between Seyfert Subtypes}
\label{sec:discussion_cross_subtype}

The controlled Random Forest experiment establishes that the two subtype specific photometric relations are not interchangeable. Because both transfer directions used the same algorithm, hyperparameters, and 45 input features, the loss of performance cannot be attributed to differences between the primary regression algorithms.

The matched control strengthens this conclusion. Equalising both sample size and spectroscopic redshift distribution did not remove the transfer degradation across 50 independently constructed realisations. The two directions nevertheless retained different error structures: transfer from Seyfert~I to Seyfert~II showed the stronger systematic underestimation and larger absolute errors, while transfer from Seyfert~II to Seyfert~I produced the broader normalised residual distribution and the larger outlier fraction. This is why the transfer directions should not be ranked from \(R^2\) alone, which depends on the variance of the evaluated redshift distribution.

The feature ablation tests provide an additional constraint on possible selection effects. Removing the broadband flux densities did not remove the transfer loss, so the result cannot be explained solely by apparent brightness information. Restricting the analysis to SDSS photometry also preserved the transfer asymmetry, while WISE only models performed substantially worse in both directions. Within the present redshift range, the optical bands therefore carry most of the predictive redshift information, with WISE providing complementary information when combined with SDSS. The persistence of the transfer loss in colours only and SDSS only experiments argues against a single photometric component being responsible for the effect.

A plausible interpretation is that the broad-line Seyfert~I sample spans a wider range of nuclear to host contributions than the more host-dominated Seyfert~II population. A model trained on the broader Type I photometric distribution may therefore retain some predictive ability for the narrower Type II domain, while a relation calibrated on Type II photometry does not represent the full range of unobscured nuclear continua. This interpretation is consistent with work showing that AGN photo-\(z\) performance can benefit from separating sources according to spectral or photometric properties before redshift estimation \citep{Salvato2009,Salvato2011,FotopoulouPaltani2018}. However, the two samples originate from different SDSS catalogues and do not have identical luminosity distributions. The observed asymmetry should therefore be interpreted as a robust property of the catalogue selections studied here rather than as a unique measurement of obscuration physics.

The redshift extrapolation test shows that the applicability domain is not limited to Seyfert subtype. When the regressors were trained only at \(z_{\mathrm{spec}} \leq 0.5\), both models systematically underestimated the redshifts of sources above this limit. The relatively small NMAD values in this test do not imply accurate extrapolation, because NMAD measures residual scatter around the median residual and is therefore less sensitive to a coherent systematic offset. The substantially increased MAE and negative bias instead show that the strong within-subtype performance obtained across the full \(0 < z_{\mathrm{spec}} \leq 0.6\) samples should not be interpreted as evidence that these models can be extrapolated beyond the redshift coverage of their training data. Representative spectroscopic training data across the intended redshift range therefore remain necessary.

The 36 pairwise colours are algebraically redundant and were used as a predictive representation for tree-based models, not as 36 independent physical observables. For this reason, we do not assign physical meaning to individual tree feature importance values. The grouped ablation tests provide a more direct assessment of the contribution of colour, flux density, optical, and mid-infrared information.

\subsection{Selective Prediction with a Seyfert Subtype Classifier}
\label{sec:discussion_selective_prediction}

The classifier was introduced to restrict the domain in which the Seyfert~II redshift model is applied. This differs from using subtype as an auxiliary label after regression because the classification probability determines whether a Seyfert~II photo-\(z\) estimate is retained. The development sample was balanced between Seyfert~I and Seyfert~II within redshift bins, and the probability threshold was selected only from out-of-fold development predictions before evaluation on the untouched test data. In that independent evaluation, the classifier retained most Seyfert~II sources while rejecting the large majority of Seyfert~I sources. The external QSO sample, which was excluded from classifier training and threshold selection, showed similarly low acceptance. The classifier therefore does not simply treat all spectroscopically confirmed AGN as appropriate inputs to the Seyfert~II regressor.

The independent gate experiment provides a second line of evidence that the selection is informative about regression reliability. Among the true Seyfert~II objects in the untouched test sample, the accepted subset had lower MAE, \(\sigma_{\mathrm{NMAD}}\), and outlier fraction than the rejected subset. The classification probability is therefore associated not only with subtype separation but also with the reliability of the subsequent specialised redshift estimate. This result supports classification assisted regression strategies used in previous photo-\(z\) studies \citep{FotopoulouPaltani2018,Li2022}, while differing from approaches in which Type I and Type II AGN are fitted jointly by one regressor and subtype is predicted only as an additional output \citep{Mechbal2024}.

The classifier is not intended for unrestricted survey catalogues. Stars, inactive galaxies, star forming galaxies, and other source populations were outside the classification problem, and the present analysis produces point photo-\(z\) estimates rather than full redshift probability distributions. Extending the method to less complete multiband samples, additional source classes, external photometric catalogues, and probabilistic redshift outputs will be necessary before survey scale application.

\section{Conclusions}
\label{sec:conclusions}

We tested whether a photometric redshift relation calibrated for one low-redshift Seyfert subtype remains applicable to the other, using a common SDSS+WISE photometric representation and controlled machine learning experiments. Our main conclusions are:

\begin{itemize}

\item The subtype specific models achieved different levels of within sample accuracy. The Seyfert~II Random Forest model reached an MAE of \(0.02008\), \(\sigma_{\mathrm{NMAD}}=0.01691\), and an outlier fraction of \(0.206\%\), while the Seyfert~I XGBoost model yielded an MAE of \(0.03852\), \(\sigma_{\mathrm{NMAD}}=0.02962\), and an outlier fraction of \(0.988\%\).

\item The subtype specific photo-\(z\) relations were not interchangeable. In the controlled Random Forest experiment, the Seyfert~II to Seyfert~I transfer produced an MAE of \(0.06819\), \(\sigma_{\mathrm{NMAD}}=0.05336\), and an outlier fraction of \(3.996\%\). The transfer degradation remained present across 50 realisations matched in sample size and spectroscopic redshift distribution.

\item The transfer loss persisted when the analysis was restricted to colours alone and when only SDSS photometry was used. The effect therefore cannot be attributed solely to apparent flux density information or to the inclusion of the WISE bands. SDSS photometry carried most of the predictive redshift information in the present low-redshift samples, while WISE provided complementary information in the combined feature set.

\item The probabilistic subtype classifier provided a useful gate for the specialised Seyfert~II regressor. At \(P(\mathrm{Sy\,II})\geq0.620\), it accepted \(89.61\%\) of the untouched Seyfert~II test sample, compared with \(4.09\%\) of the Seyfert~I test sample and \(2.48\%\) of the external QSO sample. Among the independent Seyfert~II test sources, accepted objects also showed lower photo-\(z\) error than rejected objects, with MAE values of \(0.03320\) and \(0.04986\), respectively.

\end{itemize}

The redshift extrapolation test further showed that this domain dependence also applies to redshift coverage. Models trained only at \(z_{\mathrm{spec}} \leq 0.5\) systematically underestimated sources above that training limit, demonstrating that the reported within-sample accuracy should not be interpreted as reliable extrapolation beyond the spectroscopic redshift support of the training data.

These results show that the reliability of an AGN photo-\(z\) calibration depends on the population represented in its training data. For preselected low-redshift AGN samples, subtype specific regression combined with probabilistic source selection provides a more defensible strategy than assuming that a single Seyfert calibration transfers unchanged across broad and narrow-line populations.

\section{Acknowledgements}

The author thanks his academic mentor, Dr H. Tuğça Şener, for her continued support throughout his scientific journey and for fostering his interest in data science and astrostatistics.

\section{Code Availability}

The complete analysis pipeline, feature engineering code, trained Random Forest and XGBoost models, and supporting results are available at \url{https://github.com/uzayyaydiin/seyfert-photoz}.

\bibliography{sn-bibliography}

\section*{Statements and Declarations}

\textbf{Funding}

The author declares that no funds, grants, or other support were received during the preparation of this manuscript.

\textbf{Competing Interests}

The author has no relevant financial or non-financial interests to disclose.

\textbf{Author Contributions}

Uzay Aydin conceived the study, performed the data preparation and machine-learning analyses, interpreted the results, prepared the figures and tables, and wrote the manuscript.

\textbf{Data Availability}

The spectroscopic and photometric data used in this study are publicly available through the Sloan Digital Sky Survey and the AllWISE Source Catalogue. The analysis code, feature-engineering pipeline, trained models, and supporting results are available at \url{https://github.com/uzayyaydiin/seyfert-photoz}.

\end{document}